\documentclass[a4paper]{jpconf}
\usepackage{amsmath,amssymb,bm}
\usepackage{graphicx}
\usepackage{epstopdf}
\usepackage{latexsym}
\usepackage{subfigure}
\usepackage{color}
\usepackage{hyperref}
\usepackage{braket}
\hypersetup{
  colorlinks,
  citecolor=magenta,
  linkcolor=blue,
  urlcolor=blue}
      
\newcommand{\nccp}[1]{nc-$\mathbb{CP}^{#1}$}
\newcommand{\cp}[1]{$\mathbb{CP}^{#1}$}

\begin{document}
\title{Numerical studies of various N\'eel-VBS transitions in SU($N$) anti-ferromagnets}

\author{Ribhu K. Kaul}

\address{Department of Physics \& Astronomy, University of Kentucky,
  Lexington KY 40506-0055}
\ead{rkk@pa.uky.edu}

\author{Matthew S. Block}

\address{Department of Physics \& Astronomy, California State
University, Sacramento CA 95819}
\ead{matthew.block@csus.edu}

\begin{abstract}
In this manuscript we review recent developments in the
numerical simulations of bipartite SU($N$) spin models by
quantum Monte Carlo (QMC) methods. We provide an account of 
a large family of newly discovered
sign-problem free
spin models which can be simulated in their ground states  on large lattices,
containing $O(10^5)$ spins, using the stochastic series
expansion method with efficient loop algorithms. One of the most important applications so far of these
Hamiltonians are to unbiased studies of
quantum criticality between N\'eel and valence bond phases in two
dimensions -- a summary of this body of work
is provided. The article concludes with an overview of the current status of and
outlook for future studies of the ``designer'' Hamiltonians. 
\end{abstract}

\section{Overview}

The study of ground states of lattice models of quantum spins has become a major field in condensed matter
physics~\cite{balents2010:spliq}. Despite their simplicity these
models can be
extremely hard to study theoretically, due in part to the
rich variety of ground states that they can host. 
Recent years have seen a dramatic increase in the use of
numerical methods to study the ground states of quantum spin models. Of particular interest
are ``unbiased'' numerical methods, which solve for physical properties of
model systems with numerical errors that can be controlled and
estimated in a reliable fashion. Quantum Monte Carlo occupies a special
place among the unbiased methods because when applicable, it is the only technique that
is able to access the large systems sizes required for reliable
extrapolation to the thermodynamic limit, especially in the proximity of critical phenomena~\cite{kaul2013:qmc}.

The simplest ground state that can arise in a quantum spin system is a
magnetic ground state, where the expectation value of the spin on a
site is finite, causing the spin to effectively ``point'' in a certain
direction, thus spontaneously breaking the global symmetry associated with the
rotation of spins. Such magnetic states are well known to arise in the
low-energy space of classical
spin models and their appearance in quantum models can be understood
from semi-classical arguments. 
The focus of recent numerical studies of spin models has been in large
part on the nature
of the {\em non-magnetic phases} that arise at $T=0$ due to quantum
fluctuations and the quantum phase transitions separating magnetic and
non-magnetic phases. The non-magnetic phases may either break another symmetry,
most often a lattice translational symmetry in which case they are called ``solids''
or be completely symmetric in both the lattice and spin symmetries in
which case they are called ``liquids.'' The central questions of interest
 to numerical studies are: What is the precise characterization of the non-magnetic phases, both of the ``solid''
 and ``liquid'' type that are found in simple microscopic models? How
 can the phase transitions 
between the magnetic and non-magnetic phases observed in numerical
simulations be understood in terms of long wavelength quantum field theories?

While these general questions have received copious
attention from a multitude of
complementary approaches in different contexts over the last three decades~\cite{kaul2013:qmc,Stoudenmire12,sachdev1999:qpt,laflorencie2004:ed}, in this review we will outline how 
new ``designer'' Hamiltonians with SU($N$) symmetry have contributed
to answers concerning phase transitions between SU($N$) symmetry
breaking magnetic phases and lattice translational symmetry breaking
valence-bond solid states, for which  an exotic
direct continuous transition -- ``deconfined criticality'' -- has been
proposed~\cite{senthil2004:science,senthil2004:deconf_long,senthil2005:jpsj}. For a
fuller appreciation of this short review, a familiarity with the
deconfined theory is recommended. In the interest of space this is not
provided here, the interested reader is encouraged to
refer to the original literature.
 In Section~\ref{sec:models} we briefly review the discovery of a 
large family of ``designer'' SU($N$) models that do not suffer from the sign
problem. 
In Sec~\ref{sec:pd_deconf} we describe the phase diagrams obtained
for the Hamiltonians that have been studied so far and the nature of the
critical points that arise between different phases. Finally,
in Section~\ref{sec:outlook} we conclude with an outlook and
directions for future work.

\section{SU($N$) Hamiltonians, loop models and the sign problem}
\label{sec:models}

In order to use unbiased quantum Monte-Carlo techniques efficiently, one needs to
identify models that do not suffer from the sign problem. It is well known that Hamiltonians that satisfy Marshall's sign
condition are also sign-problem free. So a basic questions of central
importance to the numerical simulations of quantum spin models is:
What is the family of spin models that satisfies Marshall's
sign criteria? 

In order to make this question concrete, we specialize our
considerations to bipartite
models with a specific representation of SU($N$) symmetry, originally
introduced to condensed matter physics by
Affleck~\cite{affleck1985:lgN}. This realization of SU($N$) symmetry
requires spins on one sub-lattice to transform in the fundamental
representation of SU($N$) and spins on the other sub-lattice to
transform in the conjugate to fundamental representation. We note here
for $N=2$, since the fundamental and conjugate to fundamental 
representations are identical, this realization of SU(2) symmetry
gives rise to
the familiar Heisenberg-like models. 

The standard quantum-classical mapping allows us to rewrite the
quantum statistical mechanics of $d$-dimensional Hamiltonians as a classical
statistical mechanics problem in $d+1$-dimensions, where the extra
dimension is of extent $\beta = 1/T$. 
The stochastic series expansion (SSE) is an elegant
and well-documented method to execute this
step~\cite{sandvik2010:vietri}. 
When carried out and as discussed in detail~\cite{kaul2014:design}, the Affleck SU($N$) spin models on bipartite lattices
can be mapped to {\em oriented tightly packed loop models with $N$ colors} in one higher
dimension. In order to carry out Monte Carlo sampling we require these configurations to have positive
weights. In the language of loop models it is straightforward to
systematically write
down all possible interactions that keep the weights of the loop configurations
positive. When the quantum-classical mapping is run backwards, the
interactions in the classical loop model correspond to terms in quantum
Hamiltonians that are Marshall positive. This picture allows
one to systematically write down a large class of SU($N$) spin Hamiltonians that are
Marshall positive. We note parenthetically that the Marshall positivity of these
``designer'' Hamiltonians is not obvious when viewed directly in the
spin language, and that the ``designer'' models include all previously
known Marshall positive spin models as particular cases. 

As a concrete example of the results, let us discuss the familiar $N=2$
case. Here the spins $\vec S$ on either sub-lattice can be written in terms
of Pauli matrices in the usual way. Previously known sign-problem-free
Hamiltonians include (with $J_1,J_2,Q>0$)~\cite{sandvik2007:deconf},
\begin{eqnarray}
\label{eq:j1j2Q}
H_{J_1} &=& J_1 {\vec S^{A}_1} \cdot {\vec S^{B}_1} \nonumber\\
H_{J_2} &=& - J_2{\vec S^{A}_1} \cdot {\vec S^{A}_2} \nonumber\\
H_{Q} &=& - Q\left ( {\vec S^{A}_1} \cdot {\vec S^{B}_1} -\frac{1}{4}\right ) \left ( {\vec S^{A}_2} \cdot {\vec S^{B}_2} -\frac{1}{4}\right )
\end{eqnarray}
where the superscript $A,B$ indicates the sublattice the spin
lives on and the subscript indicates the different spins on a given sublattice. Thus the $J_1$ interaction is defined on two spins, one of
which lives on the A sublattice and the other on the B
sublattice, the $J_2$ interaction is defined on two spins on the same
sublattice, and the $Q$ interaction is defined on four spins, two on
the A sublattice and two on the B sublattice. As an application of the new strategy
sketched above, a new linearly independent four-spin interaction can be shown to be Marshall
positive, 
\begin{eqnarray}
H_{R} &=& R\left ( {\vec S^{A}_1} \cdot {\vec S^{A}_2}
  -\frac{1}{4}\right ) \left ( {\vec S^{B}_1} \cdot {\vec S^{B}_2}
  -\frac{1}{4}\right )\nonumber \\ 
&-& R\left ( {\vec S^{A}_1} \cdot {\vec S^{B}_1} -\frac{1}{4}\right )
\left ( {\vec S^{A}_2} \cdot {\vec S^{B}_2} -\frac{1}{4}\right )
- R\left ( {\vec S^{A}_1} \cdot {\vec S^{B}_2} -\frac{1}{4}\right ) \left ( {\vec S^{A}_1} \cdot {\vec S^{B}_2} -\frac{1}{4}\right )
\end{eqnarray}
It is straightforward to prove the Marshall positivity of the above
interaction by direct evaluation of its matrix elements when $R>0$, even though its positivity is not apparent from a naive inspection of the term. 

We note following the strategy described in Ref.~\cite{kaul2014:design}, all the Marshall-positive
interactions involving an arbitrary large number of $A$ and $B$ spins
and with any $N$
(of SU($N$)) can 
be systematically written down. Also from the way the interactions are written above it is
clear that they can be written on any bipartite lattice in any
dimension, and can be made to preserve the lattice symmetry through an
appropriate summation over the entire lattice.
 
Detailed reviews and pedagogical introductions to the kind of QMC algorithms that are used to simulate the spin
models introduced in this Section may be found in~\cite{sandvik2010:vietri,kaul2013:qmc}.

\begin{figure}
\begin{center}
\includegraphics [trim=0cm 8cm 0cm 0cm, clip=true, width=40pc] {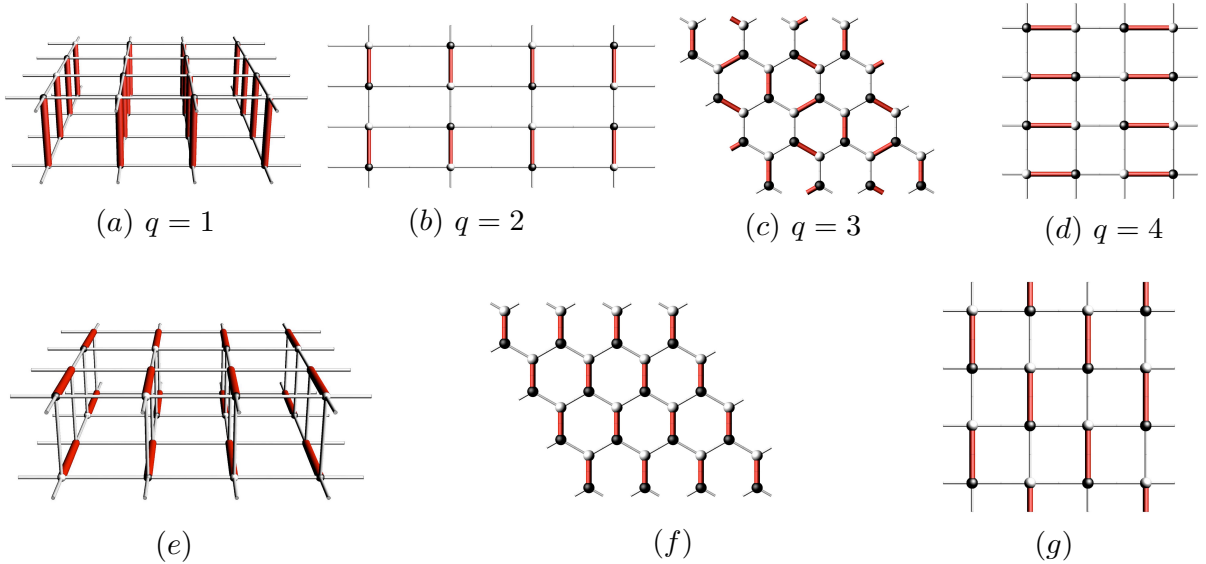}
\end{center}
\caption{\label{fig:vbsp}Summary of valence-bond patterns that have
  been accessed in numerical simulations of two-dimensional bipartite
  SU($N$) designer Hamiltonians. The quantum phase transition into each of
  these phases from SU($N$) symmetry breaking N\'eel phases is the
  subject of this review. The thick red bonds when put in an SU($N$) singlet
  state provide a cartoon state for the symmetry breaking. (a-d) are
  the ``columnar'' patterns for
  which the theory of a continuous deconfined criticality can hold for sufficiently
large $N$. The value of $q$ is the degeneracy of the VBS coverings in these states. (e) is a four-fold degenerate VBS state on a bilayer to
which there is a first order transition. (f,g) are three- and four-fold staggered VBS
states into which there are also first order transitions.}
\end{figure}

\section{Phase Diagrams \& Quantum Criticality}
\label{sec:pd_deconf}

The only two phases found so far in the numerical simulations of the bipartite SU($N$) sign-free
models discussed in Sec.~\ref{sec:models} are N\'eel and ``valence-bond''
phases (VBS). By ``valence-bond'' phases we mean phases that are smoothly
connected to a product state of two-site valence bond coverings (see
Fig.~\ref{fig:vbsp} for cartoons of such coverings). 
The numerical study of the N\'eel and
VBS phases of the Affleck SU($N$) models was initiated early on~\cite{Santoro99,harada2003:sun}. However, the
most intriguing aspect of the phase diagram, the critical point, was
first accessed with the introduction of the four-site $Q$ interaction~\ref{eq:j1j2Q}
for the $N=2$ case~\cite{sandvik2007:deconf}. While the pioneering study was carried out for
$N=2$ on the square lattice, subsequent work has carried out the study
for a large range of $N$ and for a variety of bipartite lattices and
interactions. 

In order to summarize the studies  in
``conceptual'' rather than historical order, we need one technical
result from the deconfined theory. The deconfined theory predicts that
for bipartite SU($N$) magnets in two dimensions, the N\'eel-VBS critical point is described by a
\nccp{N-1} critical field theory, only for certain ``columnar'' VBS
states. In actual lattice realization, whether in numerical
simulations or real materials, a
perturbation to the \nccp{N-1} theory, $\lambda_q$ (for the experts, it
is the fugacity of $q$-monopoles in the gauge field) is present.
The central difference between the various bipartite
lattices is the integer $q$, which can be intuitively understood as
the minimum degree of degeneracy of the VBS phase on the particular
bipartite lattice under consideration.
In order for the deconfined critical point to exist $\lambda_q$ must be {\em irrelevant} at the monopole-free fixed
point. If it is relevant one expects a first order transition or a
continuous transition in a universality class different from the
\nccp{N-1} universality. 

\begin{figure}
\begin{center}
\includegraphics [trim=0cm 0cm 0cm 0cm, clip=true, width=\columnwidth] {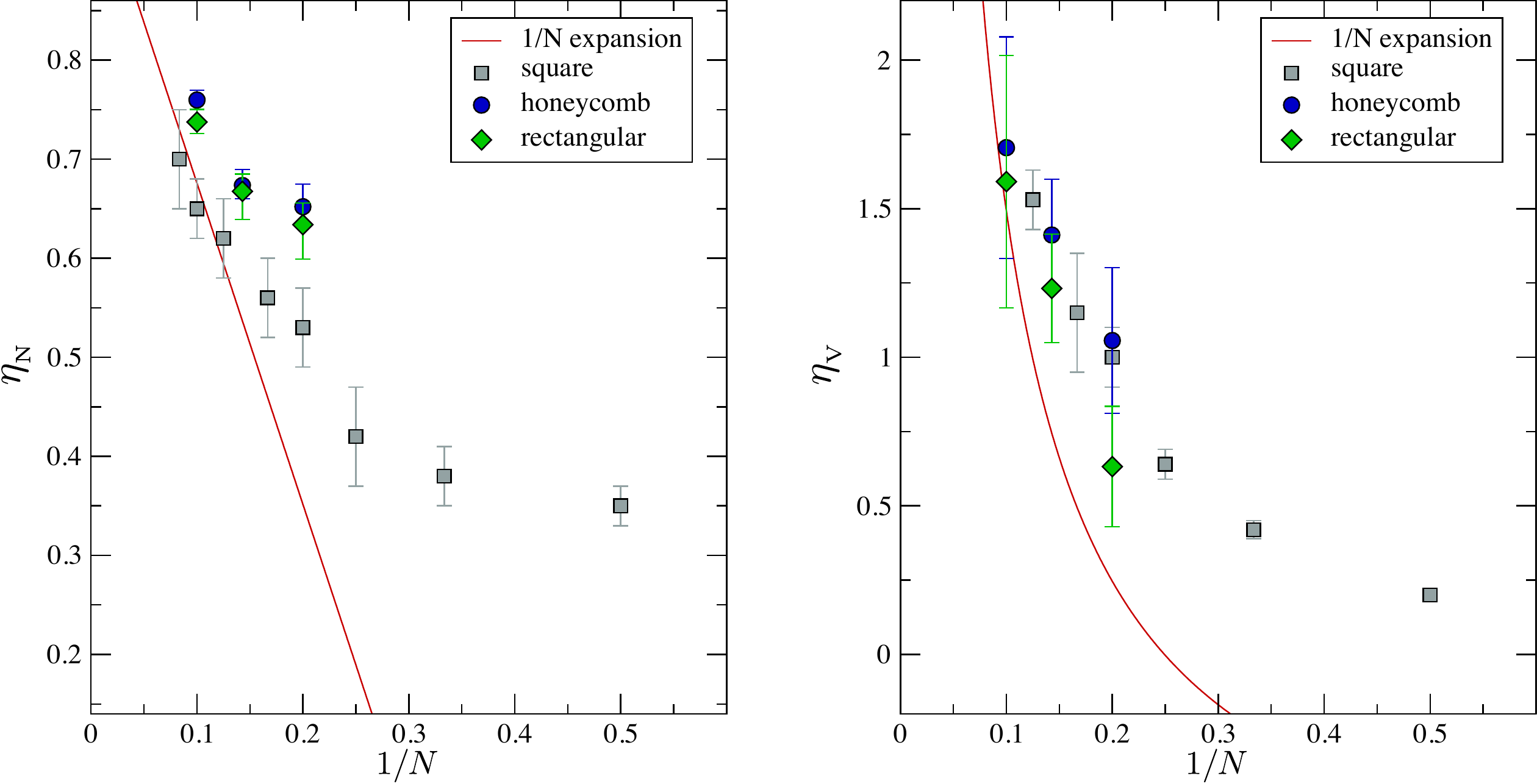}
\end{center}
\caption{\label{fig:eta}  Comparison of anomalous dimensions of N\'eel
  and VBS operators in the case of continuous transitions for $q=2,3$
  and $4$. (a) Anomalous dimension of the N\'{e}el order parameter as
  a function of $1/N$.  (b) Anomalous dimension of the VBS order
  parameter as a function of $1/N$.  The gray squares are the results
  of a square lattice study ($q=4$)
  ~\cite{lou2009:sun,kaul2012:j1j2}. The blue circles are results
  from the honeycomb lattice ($q=3$) and the green diamonds are
  results from the rectangular lattice ($q=2$).  The red line is the
  $1/N$ expansion. The universality of the exponents with respect to
  $q$ is a direct consequence of the irrelevance of $\lambda_q$. The
  agreement of the exponents
with the $1/N$ computation is strong evidence for the emergence of the
\nccp{N-1} universality at the N\'eel-VBS transition. Adapted from Ref.~\cite{block2013:fate}.}
\end{figure}

Let us begin with the case of the bilayer lattice. The simplest non-magnetic state in the bilayer system is a
non-degenerate ($q=1$) rung singlet state (see Fig.~\ref{fig:vbsp}(a)).
 For $N=2$ there is beautiful evidence that supports the theoretical expectation~\cite{chakravarty1988:qaf}
 that there is a continuous critical point
 between N\'eel and rung-singlet state in the $2+1$-dimensional O(3)
 universality class~\cite{wang2006:bilayer}. For $N\geq 4$ numerical simulations have found
 evidence for a
 first order transition, which is expected from Landau theory~\cite{kaul2012:bilayer}. The
 case $N=3$ appears to have either a very weakly first order or a
 continuous transition. The field theoretical scenario explaining a
 possible continuous $N=3$ transition has been nicely
 summarized in Ref.~\cite{nahum2013:long}.
Other numerical work has studied the N\'eel-VBS transition on the
rectangular ($q=2$)~\cite{block2013:fate}, honeycomb
($q=3$)~\cite{block2013:fate,pujari2013:hc,harada2013:deconf} and
square
($q=4$)~\cite{sandvik2007:deconf,lou2009:sun,kaul2012:j1j2,kaul2011:su34}
lattices for a range
of $N$ by using a judicious choice of the designer couplings defined
in  Section~\ref{sec:models}. See Fig.~\ref{fig:vbsp}(b-d) for
cartoons of the VBS states. Both first order transitions for small $N$
and continuous transitions for large $N$ have been identified. The
critical value of $N$ at which the transition turns continuous
decreases as $q$
increases, as expected theoretically. The appearance of
continuous and first-order transitions occur in a systematic way
that can be attributed to whether the lattice anisotropy in the form
of $\lambda_q$ is relevant
or irrelevant at the  \nccp{N-1} fixed point. The results may be
summarized in Table~\ref{tab:qN}, which shows whether the deconfined
critical point between an SU($N$) magnet and $q$-degenerate VBS state
(when it is the minimum degeneracy on a particular lattice) is
stable. We note that in the bilayer model ($q=1$) the $R$ is expected to
turn to $I$ at a finite value of $N$ -- analytic estimates from large-$N$ theories suggest this happens around $N\approx 25$~\cite{murthy1990:mono}. However, such
large values of $N$ have not
been accessed numerically yet. For the values of $N$ for which there is a
continuous transition, the critical exponents should not depend on the
bipartite lattice geometry, since $\lambda_q$ is irrelevant if there
is a continuous transition. Numerical measurements for the anomalous
dimension of the N\'eel and VBS fields are consistent with this
expectation and are consistent with results from the analytic $1/N$
expansion as shown in Fig.~\ref{fig:eta}

\begin{center}
\begin{table}[t]
\centering
\begin{tabular}{||c||c|c|c|c|c|c|c||} 
\hline 
\hline
 $N=\infty, 1/N$ & $I$ & $I$ & $I$ & $I$ & $\dots$& $I$   &  nc-$\mathbb{CP}^{N-1}$\\
\hline
\hline
$\dots$ & $$ & $$ & $$ & $$ & $$& $$  & $$\\
\hline
 $N=10$ & $R$ & $I$ & $I$ & $I$ & $$  & $I$&  nc-$\mathbb{CP}^{9}$\\
\hline
 $N=9$ & $R$ & $I$ & $I$ & $I$ & $$  & $I$&  nc-$\mathbb{CP}^{8}$\\
\hline
 $N=8$ & $R$ & $I$ & $I$ & $I$ & $$  & $I$&  nc-$\mathbb{CP}^{7}$\\
\hline
 $N=7$ & $R$ & $I$ & $I$ & $I$ & $$  & $I$&  nc-$\mathbb{CP}^{6}$\\
\hline
 $N=6$ & $R$ & $I$ & $I$ & $I$ & $$  & $I$&  nc-$\mathbb{CP}^{5}$\\
\hline
 $N=5$ & $R$ & $I$ & $I$ & $I$ & $$  & $I$&  nc-$\mathbb{CP}^{4}$\\
 \hline
 $N=4$ & $R$ & $I$ & $I$ & $I$ & $$  & $I$&  nc-$\mathbb{CP}^{3}$\\
 \hline
 $N=3$ & $R$ & $R$ & $I$ & $I$ & $$ & $I$&  nc-$\mathbb{CP}^{2}$ \\
 \hline
$N=2$ & $R$ & $R$ & $I$ & $I$ & $$ & $I$& nc-$\mathbb{CP}^{1}$\\
\hline 
$N=1$ & $R$ & $R$ & $R$& $I$ & $$& $I$ & $XY$\\
\hline
$N=0$ & $R$ & $R$ & $R$& $R$ & $$& $R$ & photon\\
\hline
$$ & $q=1$ & $q=2$ & $q=3$& $q=4$ & $\dots$ & $q=\infty$ & $$\\
\hline
\hline 
\end{tabular}
\caption{Table showing the inferred relevance ($R$) or irrelevance ($I$) of
  $q$-monopoles at the \nccp{N-1} fixed point, which various studies
  summarized in the text have allowed us to complete. Numerical simulations of the N\'eel-VBS transition in the models discussed here only allow studies for $N\geq 2$.  The entries with $R$
  correspond to an unstable fixed point, and $I$ to a stable fixed
  point allowing for deconfined criticality. At some currently unknown critical value of $N>10$, the $q=1$
  case switches from $R$ to $I$. Adapted from Ref.~\cite{block2013:fate}  }
\label{tab:qN}
\end{table}
\end{center}

Having discussed the cases of a ``columnar'' VBS state to which the
theory of deconfined criticality applies, it is also of interest to
numerically study the N\'eel-VBS transition in situations where the deconfined
criticality does {\em not} apply, as a non-trivial check on the theory. As we shall see in the examples below there are a
number of different reasons why this can happen:

(1) In the bilayer geometry, consider the phase
transition between a four-fold degenerate columnar VBS (c-VBS), see
Fig.~\ref{fig:vbsp}(e), and the SU($N$) N\'eel state. The c-VBS  on
the bilayer  breaks
exactly the same symmetries as the single-layer c-VBS state
(see Fig. ~\ref{fig:vbsp}(d)) and thus one might conclude naively
based on ``Landau'' theory that
they are described by the same critical phenomena. However, it is well
known that in the bilayer geometry the Berry phases that are crucial
for deconfined criticality cancel between the layers, eliminating the
possibility of an exotic continuous transition.  Consistent with this
expectation, numerical studies have found that the c-VBS-N\'eel transition in the bilayer is 
first order~\cite{kaul2012:bilayer}.

(2) Back to the single-layer case, one can tune designer coupling to
favor spontaneous symmetry breaking different from the ``columnar'' states for
which the original proposal of deconfined criticality was made~\cite{senthil2004:science}. 
Indeed, studies have been carried out for the transition between N\'eel and ``staggered''
VBS for SU(2), on the square lattice by tuning a ``designer'' six-spin
interaction~\cite{sen2010:first} and on the honeycomb by tuning a four-spin
interaction~\cite{banerjee2011:sthc}. The theory of a continuous deconfined critical point
does not generalize to the staggered VBS case, and in the absence of
any plausible alternative one expects a
restoration of ``Landau'' theory and hence a first-order transition. Consistently, both studies find clear
evidence for first-order transitions between the N\'eel and staggered
VBS phases.

(3) There have been two studies of the N\'eel-VBS transition in designer
Hamiltonians in {\em three- dimensional} bipartite SU($N$)
systems thus far. Both studies, which are on cubic lattices but
with distinct Hamiltonians, find first-order transitions between
N\'eel and columnar VBS and no evidence for any new intervening
phases~\cite{beach2007:cubic,block2012:cubic}. The absence of a direct continuous transition in $3+1$ dimensions is again consistent with the deconfined
criticality scenario since various aspects of the \cp{N-1} field theory
are specific to $2+1$ dimensions and are known to be invalid in $3+1$ dimensions.

\section{Outlook}
\label{sec:outlook}

In the previous section, we have summarized how the deconfined
criticality scenario and the \cp{N-1} universality, convincingly
explain unbiased numerical studies of the SU($N$) designer
Hamiltonians in a variety of studies carried out by different groups, which have
probed various distinct aspects of the
deconfined criticality scenario. 

One concern is that the transition could be weakly
first order and the numerical studies, which are necessarily limited to
finite size systems, may not have accessed sizes large enough to detect
a first-order transition~\cite{kuklov2008:first}.   If this is indeed the case, the transition
must be so weakly first order that the correlation length 
exceeds the sizes of the largest lattices on which many of the
designer Hamiltonians have been simulated, since no direct evidence for
first-order behavior has been found in the cases where a continuous
transition has been claimed. This makes the discussion of
the presence of such a weakly first-order transition somewhat
academic.

Nevertheless, there are well established corrections to scaling observed
in the numerical studies of the N\'eel-VBS transition in the designer
Hamiltonians~\cite{sandvik2010:logs,banerjee2010:log, kaul2011:su34,harada2013:deconf}. The origin of these
corrections to the asymptotic behavior is not completely understood. Three options might be considered: (a) the
transition is described completely by deconfined criticality; the
corrections to scaling are just the usual corrections to asymptotic
behavior that arise from deviants from the true fixed point due to
irrelevant operators, finite size, etc. and will vanish in the
thermodynamic limit; (b) the transition is weakly
first order and no non-compact \cp{N-1} fixed point exists; the
deviation from scaling is claimed as incipient first order behavior; (c)
there is something fundamentally new in the scaling behavior of
deconfined critical points that needs to be understood; the
``corrections'' are part of the true asymptotic behavior and their existence can be
understood from a field theoretic argument that has so far been overlooked. 

Currently, it is not possible to rule out any of the three
options categorically. Given the large body of evidence
presented in Section~\ref{sec:pd_deconf} that appears consistent with
the deconfined scenario and the lack of direct evidence for a first-order transition, Occum's razor suggests that (a) is the most likely
explanation. The corrections to scaling arising due to a irrelevant
operator with a small exponent~\cite{bartosch2013:frg}. As a
theoretical possibility, option (c) is the
most exciting and is an interesting
area for further field theoretic work~\cite{sandvik2011:spinon, nogueira2012:logs}.  Likewise, some positive theoretical
reasoning that supports the existence of a first-order transition and
explains why it is so weak, or direct evidence numerical evidence for
a first-order transition,
would strengthen the case for option (b). 

It is also
possible that the transition is first order for small-$N$ and
becomes continuous only for some finite value of $N>2$. Numerical studies
on the designer Hamiltonians do not see a dramatic difference in the
simulations between $N=2$ and $N> 2$ for the cases where a
continuous transition is found. It would be of interest to
extend the studies of the $N=2$ \nccp{N-1} field theory on a lattice,
to $N>2$, which have not been
carried out~\cite{kuklov2008:first,motrunich2008:cp1}.

Beyond the study of deconfined criticality, an as yet unanswered question is what other phases can
be accessed in the family of sign-free SU($N$) spin models presented
in Sec.~\ref{sec:models}. The
only two phases found so far in these models are N\'eel and VBS
phases, with the VBS phases being the
simplest possible,
i.e., they are each connected without phase transition to a cartoon
state that is simply a direct product of two-site SU($N$) singlets. There is no evidence for plaquette VBS states in the designer models, despite many
approximate studies favoring such a state in similar models. A related unanswered
question is whether the SU($N$) ``designer'' Hamiltonians can host a
spin liquid, and if so how to design such models. We note that Marshall positive models
with simpler U(1) symmetries have been shown convincingly to host
$Z_2$ spin liquids~\cite{isakov2011:tee,isakov2006:sl}, but no such model with SU($N$) symmetry is
known yet.

Finally, studies of the designer models with internal symmetries
different from SU($N$), studies of their phase diagrams with quenched
disorder, and studies of 
their spectral properties and doping with a small concentration of holes all provide exciting
directions for future research.

\ack

The authors would like to acknowledge Jon
D'Emidio, Roger Melko, and Anders
Sandvik for collaboration on related work and NSF DMR-1056536  for partial
financial support.

\section*{References}
\bibliography{/Users/rkk/Physics/PAPERS/BIB/career.bib,/Users/rkk/Physics/PAPERS/BIB/rev_bib.bib}

\end{document}